\documentclass[12pt]{iopart}

\begin{document}

\title{Boson Normal Ordering via Substitutions and Sheffer-type Polynomials}
\author{P Blasiak$^{\dag,\diamondsuit}$, A Horzela$^\dag$, K A Penson$^{\diamondsuit}$, G H E Duchamp$^{\ddag}$ and A I Solomon$^{\diamondsuit,\sharp}$}
\address
{$^\dag$ H.Niewodnicza\'nski Institute of
Nuclear Physics, Polish Academy of Sciences\\
ul. Eliasza-Radzikowskiego 152,  PL 31342 Krak\'ow, Poland}
\address
{$^\diamondsuit$ Laboratoire de Physique Th\'eorique des Liquides,\\
Universit\'e Pierre et Marie Curie, CNRS UMR 7600\\
Tour 24 - 2e \'et., 4 pl. Jussieu, F 75252 Paris Cedex 05, France}
\address
{$^\ddag$ Institut Galil\'ee, LIPN\\
99 Av. J.-B. Cl\'ement, F 93430 Villetaneuse, France}
\address
{$^{\sharp}$ The Open University, Physics and Astronomy Department\\
Milton Keynes MK7 6AA, United Kingdom}

\eads{\linebreak\mailto{blasiak@lptl.jussieu.fr}, \mailto{andrzej.horzela@ifj.edu.pl},
\mailto{penson@lptl.jussieu.fr},\mailto{ghed@lipn-univ.paris13.fr}, \mailto{a.i.solomon@open.ac.uk}\linebreak}

\begin{abstract}
We solve the boson normal ordering problem for $(q(a^{\dagger})a +
v(a^{\dagger}))^n$ with arbitrary functions $q$ and $v$ and
integer $n$, where $a$ and $a^\dag$ are boson annihilation and
creation operators, satisfying $[a,a^\dag]=1$. This leads to
exponential operators generalizing the shift operator and we show
that their action can be expressed in terms of substitutions. Our
solution is naturally related through the coherent state
representation to the exponential generating functions of
Sheffer-type polynomials. This in turn opens a vast arena of
combinatorial methodology which is applied to boson normal
ordering and illustrated by a few examples.
\end{abstract}

\section{Introduction}\label{introd}
Three decades ago Navon {\cite{[Navon]}} and Katriel
{\cite{[Katriel1]}} observed that the normal
 ordering  of quantum operators is a combinatorial problem.
Navon considered general products of fermion creation
$a^{\dagger}$ and annihilation $a$ operators and showed that such
monomials can be expressed as sums of lower order normal products
with coefficients which are combinatorial numbers called {\em rook
numbers}. Katriel analysed the normally ordered form of $(a^\dag
a)^n$ for the boson case $\left[a,a^{\dag}\right] = 1$ and arrived
at a solution given in terms of the {\em Stirling numbers of the
second kind}. Following these ideas we presented closed form
solutions for the normally ordered form of Taylor expandable boson
operator functions $F({\hat w}(a,a^{\dag}))$ where ${\hat
w}(a,a^{\dag})$ is either a  monomial in $a$ and $a^{\dag}$
operators or a monomial in $a + a^{\dagger}$ \cite{[BDHPS1]}.
Using methods of combinatorial analysis we  identified
combinatorial numbers
 appearing in such normally ordered expressions and described their properties.
 Recently Witschel \cite{[Witschel]} presented an alternative technique for this problem.

 \noindent In this Letter we  solve the problem of normally ordering  expressions of the form
 $(q(a^{\dagger})a + v(a^{\dagger}))^n$
 for arbitrary functions $q(a^{\dagger})$ and $ v(a^{\dagger})$.
This generalizes results \cite{[Mikhailov]},
\cite{[Katriel2]} obtained in the mid 80's.
 Our considerations have been
inspired by operational methods proposed by Dattoli {\it et al.}
\cite{[Dattoli1]}, \cite{[Dattoli2]}. Our approach connects the
boson normal ordering problem to the notion of substitution groups
and, if one employs the coherent
 state representation, to the theory of Sheffer-type polynomials appearing in various
areas of combinatorics \cite{[Roman]},\cite{[HierDob]}.

\section{Operational formulas via substitution group}\label{genform}
We consider bosons satisfying
$\left[a,a^{\dag}\right] = 1$ , and  use the
representation of the Heisenberg - Weyl algebra in terms of
multiplication and derivative operators $
\left[\frac{d}{dx}, x\right] = 1$.
We start with  the Taylor formula
 $\exp\left(\lambda\frac{d}{dx}\right)F(x) = F(x + \lambda)$
 and consider a generalization
of the shift operator of the form
$E_{q,v}(\lambda)=\exp{\left[\lambda\left(q(x)\frac{d}{dx} +
v(x)\right)\right]}$, where $q(x)$ and $v(x)$ are arbitrary
functions. We shall find a formula for $E_{q,v}(\lambda)F(x)$. In
fact, as we will demonstrate, the following equality holds
\begin{equation}
\label{general1}
\begin{array}{l}
\exp{\left[\lambda\left(q(x)\frac{d}{dx} + v(x)\right)\right]}F(x) = g(\lambda, x)
\cdot F(T(\lambda,x))
\end{array}
\end{equation}
\noindent  where
\begin{equation}
\label{general2}
\begin{array}{lcl}
\frac{\displaystyle{\partial}T(\lambda,x)}{\displaystyle{\partial}\lambda} =  q(T(\lambda,x))\ , &~~~~~~~~~~~~~~&T(0,x) = x\ ,
\end{array}
\end{equation}
\begin{equation}
\label{general3}
\begin{array}{lcl}
\frac{\displaystyle{\partial}g(\lambda,x)}{\displaystyle{\partial}\lambda} =  v(T(\lambda,x))\cdot g(\lambda,x)\ , &~~~~&g(0,x) = 1\ .
\end{array}
\end{equation}
\noindent We now exploit basic properties of Eqs.(\ref{general1})-(\ref{general3}).
First observe from Eq.(\ref{general1}) that the action of
$E_{q,v}(\lambda)$ on a function $F(x)$ amounts to:  a) a change
of argument $x\to T(\lambda,x)$ in $F(x)$ which is in fact a
substitution; b) multiplication by a prefactor $g(\lambda,x)$
which we call a {\it prefunction}. We also see from Eq.(\ref{general3})
that $g(\lambda,x) = 1$ for $v(x) = 0$. Finally,
note that $E_{q,v}(\lambda)$ with $\lambda$ real generates an
abelian, one-parameter group, implemented by Eq.(\ref{general1});
this gives  the following group composition law for $T(\lambda,x)$
and $g(\lambda,x)$:
\begin{equation}
\label{group}
\begin{array}{l}
T(\lambda + \theta,x)) = T(\theta,T(\lambda,x)),
\\
g(\lambda + \theta,x) = g(\lambda,x)\cdot g(\theta,T(\lambda,x)).
\end{array}
\end{equation}
\noindent  In order to prove Eqs.(\ref{general1})-(\ref{general3})
first recall the exponential mapping formula which for operators
$A$ and $B$ states that
\begin{equation}\label{e}
\begin{array}{l}
e^{\lambda A}B e^{-\lambda A} = B + \lambda [A,B] +
\frac{\lambda^2}{2!} [A,[A,B]] + ...\ .
\end{array}
\end{equation}
Calculating appropriate commutators one can check that for
arbitrary functions $v(x)$ and $G(x)$, where $x$ is the
multiplication operator, the following operator relation holds\footnote[1]{A word of explanation is appropriate when comparing Eqs.(\ref{general1}) and (\ref{lemma1}): Eq.(\ref{general1}) is an equality between functions, whereas Eq.(\ref{lemma1}) is an operator equation. Upon rewriting Eq.(\ref{lemma1}) as $e^{\lambda\left[\frac{{d}}{{\rm d}x} + v(x)\right]}G(x)= G(x + \lambda)e^{\lambda\left[\frac{{d}}{{\rm d}x} + v(x)\right]}$ and by acting with it on a constant function $1$ we obtain a particular case of Eq.(\ref{general1}) for $q(x)=1$ and $g(\lambda,x)\equiv e^{\lambda\left[\frac{{d}}{{d}x} + v(x)\right]}\ 1$. For the equation for $g(\lambda,x)$ see below.}
\begin{equation}
\label{lemma1}
\begin{array}{l}
e^{\lambda\left[\frac{{d}}{{\rm d}x} + v(x)\right]}G(x)
e^{-\lambda\left[\frac{{d}}{{\rm d}x} + v(x)\right]} = G(x +
\lambda).
\end{array}
\end{equation}
\noindent Next, in the l.h.s. of Eq.(\ref{general1}) let us change the variables according
to $x = {\tilde f}(y)$ subject to a constraint
\begin{equation}
\label{lemma1.1}
\begin{array}{l}
\frac{d}{dy}{\tilde f}(y) = q({\tilde f}(y)),
\end{array}
\end{equation}
\noindent which leads to
\begin{equation}
\label{lemma2}
\begin{array}{l}
e^{\lambda\left[q(x)\frac{{\rm d}}{{\rm d}x} + v(x)\right]}F(x)  =
e^{\lambda\left[\frac{{\rm d}}{{\rm d}y} + v({\tilde
f}(y))\right]} F({\tilde f}(y)).
\end{array}
\end{equation}
We treat the r.h.s of above equation as an operator acting on a
constant function $1$ and we get
\begin{equation}
\begin{array}{l}
e^{\lambda\left[q(x)\frac{{\rm d}}{{\rm d}x} + v(x)\right]}F(x)=
e^{\lambda\left[\frac{{\rm d}}{{\rm d}y} + v({\tilde
f}(y))\right]} F({\tilde f}(y))e^{-\lambda\left[\frac{{\rm
d}}{{\rm d}y} + v({\tilde f}(y))\right]}e^{\lambda\left[\frac{{\rm
d}}{{\rm d}y} + v({\tilde f}(y))\right]}\ 1
\end{array}
\end{equation}
\noindent which using Eq.(\ref{lemma1}) gives
\begin{equation}
\label{lemma3}
\begin{array}{l}
e^{\lambda\left[\frac{{\rm d}}{{\rm d}x} + v(x)\right]}F(x)  =
F({\tilde f}(y+\lambda)){\tilde g}(\lambda, y)
\end{array}
\end{equation}
\noindent where
\begin{equation}
\label{lemma4}
\begin{array}{l}
{\tilde g}(\lambda, y) = e^{\lambda\left[\frac{{\rm d}}{{\rm d}x}
+ v({\tilde f}(y))\right]}\ 1\ .
\end{array}
\end{equation}
\noindent  We can find the differential equation satisfied by ${\tilde g}(\lambda, y)$
\begin{equation}
\label{lemma5}
\begin{array}{lcl}
\frac{\displaystyle{\partial}{\tilde g}(\lambda, y)}{\displaystyle \partial\lambda} &=& e^{\lambda\left[\frac{{\rm d}}{{\rm d}x} + v({\tilde f}(y))\right]}v({\tilde f}(y))\ 1
\\\\
&=& e^{\lambda\left[\frac{{\rm d}}{{\rm d}x} + v({\tilde
f}(y))\right]}v({\tilde f}(y))e^{-\lambda\left[\frac{{\rm d}}{{\rm
d}x} + v({\tilde f}(y))\right]}e^{\lambda\left[\frac{{\rm d}}{{\rm
d}x} + v({\tilde f}(y))\right]}\ 1
\\\\
&=& v({\tilde f}(y) + \lambda){\tilde g}(\lambda, y),
\end{array}
\end{equation}
\noindent with the use of Eq.(\ref{lemma1}) and definition Eq.(\ref{lemma4}).
The initial condition is ${\tilde g}(0,y) = 1$.

\noindent If we identify
\begin{equation}
\label{lemma6}
\begin{array}{l}
T(\lambda,x) = {\tilde f}({\tilde f }^{-1}(x) + \lambda)
\\
g(\lambda,x) = {\tilde g}(\lambda, {\tilde f}^{-1}(x))
\end{array}
\end{equation}
\noindent then Eqs.(\ref{lemma1.1}), (\ref{lemma3}) and (\ref{lemma5}) give
Eqs.(\ref{general1})-(\ref{general3}) and so complete the proof.
Note that formulas of the type Eq.(\ref{lemma6}) were used by G.A.
Goldin in his investigations of current algebras \cite{[Goldin]}
and more recently by G. Dattoli {\it et al} \cite{[Dattoli2]}.
For a geometrical interpretation of Eq.(\ref{lemma6}) and the existence of the prefunction see also \cite{[BDHPS2]}.

Here we list several applications of
Eqs.(\ref{general1})-(\ref{general3})  for some choices of $q(x)$
and $v(x)$. Since Eqs.(\ref{general2}) and (\ref{general3}) are
first order linear differential equations we shall simply write
down their solutions without dwelling on details. First we treat
the case of $v(x) = 0$, which implies $g(\lambda,x)\equiv 1$:
\begin{equation}
\begin{array}{rclcl}
{\rm Ex.1:}~~&~~~&q(x) = x,&~~~&T(\lambda,x) =
xe^{\lambda}\nonumber
\end{array}
\end{equation}
\noindent which gives $\exp\left(\lambda x \frac{d}{{d}x}\right)F(x) = F(xe^{\lambda})$,
a well known illustration of the Euler dilation operator $\exp
\left(\lambda x\frac{d}{{d}x}\right)$.
\begin{equation}
\label{ex2}
\begin{array}{rclcl}
{\rm Ex.2:}~~&~~~&q(x) = x^{r},~~~r>1,&~&T(\lambda,x) =
\frac{\displaystyle x}{\displaystyle\left(1 - \lambda\left(r - 1\right)x^{r-1}\right)^{\frac{1}{r-1}}}\nonumber
\end{array}
\end{equation}
\noindent The above examples were already considered
in the literature \cite{[BDHPS1]}, \cite{[Dattoli1]},
\cite{[Lang]}, \cite{[BDHPS2]}.

\noindent We shall go on to examples of $v(x)\neq 0$ leading to nontrivial prefunctions:
\begin{equation}
\begin{array}{rcl}
{\rm Ex.3:}~~&~~~&q(x) = 1; ~~~~~v(x)~-~{\rm arbitrary},\nonumber
\\
&~~~&T(\lambda,x) = x + \lambda,\nonumber
\\
&~~~&g(\lambda,x) = \exp{\left[\int\limits_{0}^{\lambda}{\rm d}u\
v(x+u)\right]},\nonumber
\end{array}
\end{equation}
\begin{equation}
\begin{array}{rcl}
{\rm Ex.4:}~~&~~~&q(x) = x; ~~~v(x) = x^2;\nonumber
\\
&~~~&T(\lambda,x) = xe^{\lambda},\nonumber
\\
&~~~&g(\lambda,x) = \exp{\left[\frac{x^{2}}{
2}
\left(e^{2\lambda} - 1\right)\right]},\nonumber
\end{array}
\end{equation}
\begin{equation}
\begin{array}{rcl}
{\rm Ex.5:}~~&~~~&q(x) = x^{r},~~~r>1,~~~v(x) = x^s;\nonumber
\\
&~~~&T(\lambda,x) = \frac{\displaystyle x}{\displaystyle\left(1 -
\lambda\left(r - 1\right)x^{r-1}\right)^{\frac{1}{r-1}}},\nonumber
\\
&~~~&g(\lambda,x) = \exp{\left[\frac{\displaystyle
x^{s-r+1}}{\displaystyle 1 - r}\left(\frac{\displaystyle
1}{\displaystyle\left(1 - \lambda\left(r -
1\right)x^{r-1}\right)^{\frac{s-r+1}{r-1}}} -
1\right)\right]},\nonumber
\end{array}
\end{equation}
\noindent Closer look at above examples (or at any other example which the reader could
easily construct) indicates that the group property
Eq.(\ref{group}) may be true only if $\lambda$ satisfies certain
restrictions arising from the requirement that $T(\lambda,x)$ and
$g(\lambda,x)$ remain real-valued functions. Evidently any result
obtained with Eq.(\ref{general1})-(\ref{general3}) should be
examined in this respect and resulting restrictions on $\lambda$
be kept in mind. In general we can say that the group property
Eq.(\ref{group}) may be valid only locally
\cite{[Bourbaki]},\cite{[BDHPS2]}.

\section{Implications for boson normal ordering}\label{bosonord}
The results elaborated above will be used now to treat the problem
of the normal ordering of operator functions of boson operators.
For a general function $F(a,a^{\dag})$ its normally ordered form
${\cal N}\left[F(a,a^{\dag})\right]\equiv F(a,a^{\dag})$ is
obtained by moving all the annihilation operators $a$ to the
right, using the commutation relations. We may additionally define
the operation  $:\!G(a,a^{\dag})\!:$ which means normally order
$G(a,a^{\dag})$ {\it without} taking into account the commutation
relations. Using the latter operation
 the normal ordering problem is solved for $F(a,a^{\dag})$ if we
 are able to find an operator $G(a,a^{\dag})$ for which $F(a,a^{\dag}) = :\!G(a,a^{\dag})\!:$ is satisfied.
 To obtain the normally ordered form of functions of boson operators is a nontrivial task even for simple
functions. But it is nevertheless worthwhile, because the
knowledge of ${\cal N}\left[F(a,a^{\dag})\right]$ is very useful
in practice. It allows one to evaluate easily the expectation
values of $F(a,a^{\dag})$ in such important classes of states as
the vacuum and coherent states, the latter defined for a complex
number $z$ as $|z\rangle =
\exp{(-\frac{|z|^2}{2})}\sum_{n=0}^{\infty}\frac{z^n}{\sqrt{n!}}|n\rangle$,
(with $a^{\dag}a|n\rangle = n|n\rangle$, $\langle
n|n^{\prime}\rangle = \delta_{n,n^{\prime}}$), and satisfying
$a|z\rangle = z|z\rangle$.

Here we will apply the results of Section \ref{genform} and obtain
${\cal N}\left[e^{\lambda\left[q(a^{\dag})a +
v(a^{\dag})\right]}\right]$ for
 arbitrary functions $q(x)$ and $v(x)$. First  note that
\begin{equation}
\label{expansion1}
\begin{array}{l}
\left[q(a^{\dag})a + v(a^{\dag})\right]^{n} = h_{n}(a^{\dag})+\sum\limits_{k=1}^{n}f_{nk}(a^{\dag})a^k
\end{array}
\end{equation}
\noindent with  appropriately defined $f_{n,k}$ and $h_n$. Consequently one obtains
\begin{equation}
\label{expansion2}
\begin{array}{l}
e^{\lambda\left[q(a^{\dag})a + v(a^{\dag})\right]} = 1 +
\sum\limits_{n=1}^{\infty}\frac{\displaystyle\lambda^n}{\displaystyle
n!}\left(h_{n}(a^{\dag})+\sum\limits_{k=1}^{n}f_{nk}(a^{\dag})a^k
\right),
\end{array}
\end{equation}
\noindent whose matrix elements between the coherent states $|z\rangle$ and $|z'\rangle$
are
\begin{equation}
\label{matrixelem}
\begin{array}{l}
\langle z^{\prime}|e^{\lambda\left[q(a^{\dag})a + v(a^{\dag})\right]}|z\rangle = \langle z^{\prime}|z\rangle \left(1 + \sum\limits_{n=1}^{\infty}\frac{\displaystyle\lambda^n}
{\displaystyle n!}\left(h_{n}(z^{\prime
*})+\sum\limits_{k=1}^{n}f_{nk}(z^{\prime *})z^k\right)\right).
\end{array}
\end{equation}
\noindent The next objective is to evaluate the sum on the r.h.s. of Eq.(\ref{matrixelem}).
We do so using the representation $a^{\dag}\to x$
and $a\to \frac{d}{dx}$ and rewrite Eq.(\ref{expansion2}) as
\begin{equation}
\label{expansion3}
\begin{array}{l}
e^{\lambda\left[q(x)\frac{{d}}{{d}x} + v(x)\right]}= 1 +
\sum\limits_{n=1}^{\infty}\frac{\displaystyle
\lambda^n}{\displaystyle
n!}\left(h_{n}(x)+\sum\limits_{k=1}^{n}f_{nk}(x)\frac{\displaystyle{d}^k} {\displaystyle{d}x^k}\right).
\end{array}
\end{equation}
\noindent Next, by acting with this operator identity on $e^{yx}$ we have
\begin{equation}
\label{expansion4}
\begin{array}{l}
e^{\lambda\left[q(x)\frac{{\rm d}}{{\rm d}x} + v(x)\right]}e^{yx}
= \left[1 +
\sum\limits_{n=1}^{\infty}\frac{\displaystyle\lambda^n}{\displaystyle
n!}\left(h_{n}(x)+\sum\limits_{k=1}^{n}f_{nk}(x)y^k
\right)\right]e^{yx}.
\end{array}
\end{equation}
\noindent Eq.(\ref{general1}) allows one to rewrite the l.h.s. of the above equation (with $T(\lambda,x)$ and $g(\lambda,x)$ solutions of Eqs.(\ref{general2}) and (\ref{general3})) as
\begin{equation}
\label{finalformula1}
\begin{array}{l}
e^{\lambda\left[q(x)\frac{{d}}{{d}x} + v(x)\right]}e^{yx}
= g(\lambda,x)e^{yT(\lambda,x)},
\end{array}
\end{equation}
\noindent which, if put into Eq.(\ref{expansion4}), implies
\begin{equation}
\label{finalformula2}
\begin{array}{l}
g(\lambda,x)e^{y[T(\lambda,x) - x]} = \left[1 +
\sum\limits_{n=1}^{\infty}\frac{\displaystyle\lambda^n}{\displaystyle
n!}\left(h_{n}(x)+\sum\limits_{k=1}^{n}f_{nk}(x)y^k\right)\right].
\end{array}
\end{equation}
\noindent Using the last equality on the r.h.s of (\ref{matrixelem})
we obtain
\begin{equation}
\label{finalformula3}
\begin{array}{l}
\langle z^{\prime}|e^{\lambda\left[q(a^{\dag})a + v(a^{\dag})\right]}|z\rangle = \langle z^{\prime}|z\rangle\ g(\lambda,z^{\prime *})e^{[T(\lambda,z^{\prime *}) - z^{\prime *}]z}.
\end{array}
\end{equation}
\noindent We now apply  the crucial property of the coherent state representation
(see \cite {[Klauder]} and \cite {[Louisell]}). This is that if
for an  arbitrary operator ${F}(a,a^{\dag})$ we have
\begin{equation}
\label{basicformula1}
\begin{array}{l}
\langle z'|{F}(a,a^{\dag})|z\rangle = \langle z^\prime|z\rangle\ G(z^{\prime *},z)
\end{array}
\end{equation}
\noindent then the normally ordered form of ${F}(a,a^{\dag})$ is given by
\begin{equation}
\label{basicformula2}
\begin{array}{l}
{\cal N}\left[{F}(a,a^{\dag})\right] = :G(a^{\dag},a):\, .
\end{array}
\end{equation}
\noindent Eqs.(\ref{finalformula3}) and (\ref{basicformula2}) then provide the central result
\begin{equation}
\label{central1}
\begin{array}{l}
{\cal N}\left[e^{\lambda\left[q(a^{\dag})a +
v(a^{\dag})\right]}\right] =
:g(\lambda,a^{\dag})e^{\left[T(\lambda,a^{\dag}) -
a^{\dag}\right]a}:\ ,
\end{array}
\end{equation}
\noindent being an operator identity in which the functions $g$, $T$ are found through Eqs.(\ref{general2}) and (\ref{general3}).

\section{Normal ordering and Sheffer-type polynomials}
We now  investigate in more detail the properties of
Eq.(\ref{finalformula3}). In particular we shall identify its
inherent  characteristic polynomial structures. Using the
representation of coherent states in terms of the Glauber
displacement operator $D(z)$
\begin{equation}
\label{glauber}
|z\rangle = D(z)|0\rangle = \exp{(za^{\dagger} - z^{*}a})|0\rangle
= e^{-|z|^2/2}e^{za^{\dagger}}|0\rangle,
\end{equation}
\noindent where $|0\rangle$ is the Fock vacuum, we rewrite the l.h.s of Eq.(\ref{finalformula3}) as
\begin{equation}
\label{altgenfun1}
\begin{array}{l}
\langle z^{\prime}|e^{\lambda\left[q(a^{\dag})a + v(a^{\dag})\right]}|z\rangle =\\
=e^{-1/2\left(|z^{\prime}|^2 + |z|^2\right)}\langle 0|e^{z^{\prime
*}a}e^{za^{\dagger}}e^{-za^{\dagger}}e^{\lambda\left[q(a^{\dag})a
+ v(a^{\dag})\right]} e^{za^{\dagger}}|0\rangle.
\end{array}
\end{equation}
\noindent Next we apply the exponential mapping formula Eq.(\ref{e}) in order to get
\begin{equation}
\label{hausdorff1}
\begin{array}{l}
e^{-za^{\dagger}}e^{\lambda\left[q(a^{\dag})a +
v(a^{\dag})\right]} e^{za^{\dagger}} =
e^{\lambda\left[q(a^{\dag})(a + z) + v(a^{\dag})\right]},
\\
e^{z^{\prime *}a}e^{za^{\dagger}} = e^{za^{\dagger}}e^{z^{\prime
*}a}e^{z^{\prime *}z}
\end{array}
\end{equation}
\noindent which, if put into Eq. (\ref{altgenfun1}), give (with $\langle z^{'}|z\rangle =\exp (z^{'*}z-\frac{1}{2}|z^{'}|^2-\frac{1}{2}|z|^2)$)
\begin{equation}
\label{polynomialtype}
\begin{array}{lcr}
\langle z^{\prime}|e^{\lambda\left[q(a^{\dag})a + v(a^{\dag})\right]}|z\rangle =
\langle z^{\prime}|z\rangle e^{\frac{1}{2}|z^{\prime}|^2}\langle z^{\prime}|e^{\lambda\left[q(a^{\dag})(a + z) + v(a^{\dag})\right]}|0\rangle.
\end{array}
\end{equation}
\noindent If we expand the exponential on the r.h.s. of Eq.(\ref{polynomialtype})
 as a Taylor series in $\lambda$, then the expansion coefficients for any fixed $z^{\prime}$ are polynomials in $z$.

Properties of such polynomials may be clarified when one recalls
the definition of the family of Sheffer-type polynomials $S_n(z)$
\cite{[Roman]}. The latter are defined through the {\em exponential
generating function} (egf) as
\begin{eqnarray}\label{egfsheffer}
1+\sum_{n=1}^\infty S_n(z)\frac{\lambda^n}{n!}=A(\lambda)\
e^{zB(\lambda)}
\end{eqnarray}
where functions $A(\lambda)$ and $B(\lambda)$ satisfy: $A(0)=1$
and $B(0)=0$, $B'(0)\neq 0$. Many properties of these polynomials
may be  elucidated by means of the so called {\em umbral calculus}
\cite{[Roman]},
\cite{[Rota]}, \cite{[Gessel]} which provides us with  numerous
interesting applications.

Returning to normal order, recall that the coherent state
expectation value of Eq.(\ref{central1}) is given by
Eq.(\ref{finalformula3}). When one
\underline{fixes} $z'$ and takes $\lambda$ and $z$ as indeterminates, then the r.h.s. of Eq.(\ref{finalformula3})
may be read off as an egf of Sheffer-type polynomials defined by
Eq.(\ref{egfsheffer}). The correspondence is given by
\begin{eqnarray}
\label{AB1}
A(\lambda)=g(\lambda,z'^*),\\
\label{AB2}
B(\lambda)=\left[T(\lambda,z'^*)-z'^*\right].
\end{eqnarray}
This allows us to make the statement that the coherent state
expectation value $\langle z'|...|z\rangle$ of the operator
$\exp\left[\lambda(q(a^\dag)a+v(a^\dag))\right]$ for any fixed
$z'$ yields (up to the overlapping factor $\langle z'|z\rangle$)
the egf of a certain sequence of Sheffer-type polynomials in the
variable $z$ given by Eqs.(\ref{AB1}) and (\ref{AB2}). The above
construction establishes the connection between the coherent state
representation of the operator
$\exp\left[\lambda(q(a^\dag)a+v(a^\dag))\right]$ and a family of
Sheffer-type polynomials $S^{q,v}_n(z)$ related to $q$ and $v$
through
\begin{eqnarray}
\label{shefferseq}
\langle z^{\prime}|e^{\lambda\left[q(a^{\dag})a + v(a^{\dag})\right]}|z\rangle =
\langle z^{\prime}|z\rangle\left( 1+\sum_{n=1}^\infty S_n^{(q,v)}(z)\frac{\lambda^n}{n!}\right),
\end{eqnarray}
where explicitly (again for $z'$ fixed):
\begin{equation}
\label{shefferseq2}
\begin{array}{rcl}
S_n^{(q,v)}(z)&=&\langle z^{\prime}|z\rangle^{-1}\langle z^{\prime}|\left[q(a^{\dag})a + v(a^{\dag})\right]^n|z\rangle\\
&=&e^{\frac{1}{2}|z^{\prime}|^2}\langle
z^{\prime}|\left[q(a^{\dag})(a + z) +
v(a^{\dag})\right]^{n}|0\rangle.
\end{array}
\end{equation}
\noindent We observe that Eq.(\ref{shefferseq2}) is an extension of the seminal formula of Katriel
\cite{[Katriel1]},\cite{[Katriel2]} where $v(x)=0$ and $q(x)=x$. The Sheffer-type polynomials are
in this case exponential polynomials \cite{[Roman]} expressible
through the Stirling numbers of the second kind.

Having established relations leading from the normal ordering
problem to Sheffer-type polynomials we may consider the reverse
approach. Indeed, it turns out that for any Sheffer-type sequence
generated by $A(\lambda)$ and $B(\lambda)$ one can find functions
$q(x)$ and $v(x)$ such that the coherent state expectation value
$\langle
z'|\exp\left[\lambda(q(a^\dag)a+v(a^\dag))\right]|z\rangle$
results in a corresponding egf of Eq.(\ref{egfsheffer}) in
indeterminates $z$ and $\lambda$ (up to the overlapping factor
$\langle z'|z\rangle$ and $z'$ fixed). Appropriate formulas can be
derived from Eqs.(\ref{AB1}) and (\ref{AB2}) by substitution into
Eqs.(\ref{general2}) and (\ref{general3}):
\begin{eqnarray}\label{1}
q(x)&=&B'(B^{-1}(x-z'^*)),\\\label{2} v(x)&=&\frac{A'(B^{-1
}(x-z'^*))}{A(B^{-1 }(x-z'^*))}.
\end{eqnarray}
\noindent One can check that this choice of $q(x)$ and $v(x)$, if inserted into
Eqs. (\ref{general2}) and (\ref{general3}), results in
\begin{eqnarray}\label{3}
T(\lambda,x)&=&B(\lambda+B^{-1 }(x-z'^*))+z'^*,\\\label{4}
g(\lambda,x)&=&\frac{A(\lambda+B^{-1}(x-z'^*))}{A(B^{-1}(x-z'^*))},
\end{eqnarray}
which reproduce
\begin{eqnarray}
\langle z^{\prime}|e^{\lambda\left[q(a^{\dag})a + v(a^{\dag})\right]}|z\rangle = \langle z^{\prime}|z\rangle A(\lambda)e^{zB(\lambda)}.
\end{eqnarray}
\noindent The result summarized in Eqs.(\ref{AB1}) and (\ref{AB2}) and in their 'dual' forms Eqs.(\ref{1})-(\ref{4}),
provide us with a considerable flexibility in conceiving and
analyzing a large number of examples.

\section{Combinatorial structures}
In this section we will work out examples illustrating how the egf
($=\sum_{n=0}^\infty a(n)\frac{x^n}{n!}$) of certain combinatorial
sequences $a(n),  \; \; \; n=0,1,2,\dots$  appear naturally in the
context of boson normal ordering. To this end we shall assume
specific forms of $q(x)$ and $v(x)$ thus specifying the operator
that we exponentiate. We then give solutions to
Eqs.(\ref{general2}) and (\ref{general3}) and subsequently through
Eqs.(\ref{AB1}) and (\ref{AB2}) we shall write the egf of
combinatorial sequences whose interpretation will be given.

a) Choose $q(x)=x^r$, $r>1$ (integer), $v(x)=0$ (which implies
$g(\lambda,x)=1$). Then (see Eq.(\ref{ex2})) $T(\lambda,x) =
x\left[1 - \lambda(r - 1)x^{r-1}\right]^{\frac{1}{1-r}}$. This
gives
\begin{eqnarray}
{\cal N}\left[e^{\lambda (a^\dag)^ra}\right] =
\ : \exp\left[\left(\frac{a^\dag}{\left(1 - \lambda(r - 1)(a^\dag)^{r-1}\right)^{\frac{1}{r-1}}}-1\right)a\right]:\
\end{eqnarray}
as the normally ordered form.
We now take $z^{'}=1$ in Eqs.(\ref{general2}) and (\ref{general3})
and from Eq.(\ref{finalformula3}) one has
\begin{eqnarray}
\langle 1|z\rangle^{-1}\langle 1|e^{\lambda (a^\dag)^ra}|z\rangle  =\ \exp\left[z\left(\frac{1}{\left(1 - \lambda(r - 1)\right)^{\frac{1}{r-1}}}-1\right)\right]\ ,
\end{eqnarray}
which for $z=1$ generates the following sequences:
\begin{eqnarray}
r=2:\ \ \ \ \ \ a(n)=1,1,3,13,73,501,451,...\\
r=3:\ \ \ \ \ \ a(n)=1,1,4,25,211,2236,28471,...\ \ \ \ \ \ \ \
,\ {\rm etc.}
\end{eqnarray}
These sequences are enumerating $r$-ary forests \cite{[BDHPS1]},
\cite{[Stanley]}, \cite{[Sloane]}, \cite{[Flajolet]}.

b) For $q(x)=x\ln(ex)$ and $v(x)=0$ (implying $g(\lambda,x)=1$)
which give $T(\lambda,x)=e^{e^\lambda-1}x^{e^\lambda }$, this
corresponds to
\begin{eqnarray}
{\cal N}\left[e^{\lambda a^\dag\ln(ea^\dag)a}\right] =\ :
\exp\left[\left(e^{e^\lambda-1}(a^\dag)^{e^\lambda}-1\right)a\right]:\
,
\end{eqnarray}
whose coherent state matrix element with $z^{'}=1$ is equal to
\begin{eqnarray}
\langle 1|z\rangle^{-1}\langle 1|e^{\lambda a^\dag\ln(ea^\dag)a}|z\rangle  = \exp\left[z\left(e^{e^\lambda-1}-1\right)\right]\ ,
\end{eqnarray}
which for $z=1$ generates $a(n)=1,1,3,12,60,385,2471,...$
corresponding to partitions of partitions \cite{[Stanley]},
\cite{[Sloane]}, \cite{[Flajolet]}.

The following two examples will refer to the reverse procedure,
see Eqs.(\ref{1})-(\ref{4}). We choose first a Sheffer-type egf
and deduce $q(x)$ and $v(x)$ associated with it.

c) $A(\lambda)=\frac{1}{1-\lambda}$, $B(\lambda)=\lambda$, see
Eq.(\ref{egfsheffer}). This egf for $z=1$ counts the number of
arrangements
$a(n)=n!\sum_{k=0}^n\frac{1}{k!}=1,2,5,65,326,1957,...$ of the set
of $n$ elements \cite{[Comtet]}. The solutions of Eqs.(\ref{1})
and (\ref{2}) are: $q(x)=1$ and $v(x)=\frac{1}{2-x}$. In terms of
bosons it corresponds to
\begin{eqnarray}\label{11}
{\cal N}\left[e^{\lambda \left(a+\frac{1}{2-a^\dag}\right)}\right]
=\ :\frac{2-a^\dag}{2-a^\dag-\lambda}e^{\lambda a}:\
=\frac{2-a^\dag}{2-a^\dag-\lambda}e^{\lambda a}.
\end{eqnarray}

d) For $A(\lambda)=1$ and $B(\lambda)=1-\sqrt{1-2\lambda}$ one
gets the egf of the Bessel polynomials \cite{[Bessel]}. For $z=1$ they
enumerate  special paths on a lattice \cite{[Pittman]}.
The corresponding sequence is $a(n)=1,1,7,37,266,2431,...\ $ . The
solutions of Eqs.(\ref{1}) and (\ref{2}) are: $q(x)=\frac{1}{2-x}$
and $v(x)=0$. It corresponds to
\begin{eqnarray}\label{22}
{\cal N}\left[e^{\lambda \frac{1}{2-a^\dag}a}\right] =\
:e^{\left(1-\sqrt{(2-a^\dag)-2\lambda}\right)a}:\
\end{eqnarray}
in the boson formalism.

These examples show that any combinatorial structure which can be
described by a Sheffer-type egf can be cast in  boson language.
This gives  rise to a large number of formulas of the type
Eqs.(\ref{11}) and (\ref{22}) which are important for physical
applications. Many other examples can be worked out in detail and
will be listed in a subsequent publication.

\section{Conclusions}

The formula of Eq.(\ref{central1}) is the key result of the
present investigation. It allows one to obtain normal ordering for
a vast range of problems involving one annihilation operator (or
one creation operator through hermitian conjugation). We have
exploited the connection between the Eq.(\ref{central1}) and the
egf of Sheffer-type polynomials and have given examples of
combinatorial structures described by particular choices of $q(x)$
and $v(x)$.

\vspace{3mm}

\noindent We thank G.A. Goldin for important discussions.

\vspace{3mm}

\noindent PB wishes to thank the Polish Ministry of Scientific Research and Information Technology for support under Grant no: 1P03B 051 26.

\end{document}